\documentclass[12pt,twoside]{article}
\usepackage{fleqn,espcrc1}

\usepackage{epsfig}

\newcommand{\AmS}{{\protect\the\textfont2
  A\kern-.1667em\lower.5ex\hbox{M}\kern-.125emS}}

\title{Probing the effectiveness: chiral perturbation theory 
calculations of low-energy reactions on the deuteron}

\author{Daniel R. Phillips\address{Department of Physics,
        University of Washington, Box 351560, Seattle, WA 98195-1560}}
       
\begin{document}

\maketitle

\begin{abstract}
I provide a brief, and therefore necessarily biased, review of the 
application of effective field theory to electromagnetic processes
on the deuteron. Electron-deuteron scattering is discussed for 
virtual photon momenta up to 700 MeV, and photon-deuteron scattering
is examined for incident photon energies of order the pion mass.
\end{abstract}

\section{INTRODUCTION}

Effective field theory (EFT) is a technique commonly used in particle
physics to deal with problems involving widely-separated energy
scales. It facilitates the systematic separation of the effects of
high-energy physics from those of low-energy physics. In
strong-interaction physics the low-energy effective theory is chiral
perturbation theory ($\chi$PT)~\cite{Br95}. Here the low-energy
physics is that of nucleons and pions interacting with each other in a
way that respects the spontaneously-broken approximate chiral symmetry
of QCD. Higher-energy effects of QCD appear in $\chi$PT as
non-renormalizable contact operators. In this way the EFT provides a
long-wavelength approximation to the amplitudes for various hadronic
processes. It yields amplitudes which can be thought
of as expansions in the ratio of nucleon or probe momenta (denoted
here by $p$ and $q$) and the pion mass to the scale of chiral-symmetry
breaking, $\Lambda_{\chi \rm{SB}}$. $\Lambda_{\chi \rm{SB}}$ is
of order $m_\rho$, and so such an expansion in
$Q$, with

\begin{equation} 
Q \equiv \frac{p}{\Lambda_{\chi \rm{SB}}},
\frac{q}{\Lambda_{\chi \rm{SB}}}, \frac{m_\pi}{\Lambda_{\chi \rm{SB}}}
\label{eq:Q}
\end{equation} 
provides a controlled way to calculate hadronic processes, as long as
$p$ and $q$ are not too large. In light nuclei the typical momentum of
a bound nucleon is of order $m_\pi$ or less, and so we should be able
to calculate the response of such nuclei to low-energy probes using
$\chi$PT, thereby providing systematically-improvable,
model-independent descriptions of these reactions. In this paper I
will describe some recent work in this direction. Section 2
outlines the expansion which $\chi$PT gives for these processes, while
Sections 3 and 4 sketch this expansion's use for electron-deuteron
scattering and Compton scattering on the deuteron.

\section{POWER-COUNTING AND WAVE FUNCTIONS}

Consider an elastic scattering process on the deuteron whose
amplitude we wish to compute~\footnote{The ideas presented in this
section can easily be extended to inelastic processes, or to other
light nuclei.}.  If $\hat{O}$ is the transition operator for this
process then the amplitude in question is simply $\langle \psi| \hat{O}
|\psi \rangle$, with $|\psi \rangle$ the deuteron wave function. In
this section, we follow Weinberg~\cite{We92}, and divide the 
formulation of a systematic expansion for this amplitude into two parts:
the expansion for $\hat{O}$, and the construction of $|\psi \rangle$.

Chiral perturbation theory gives a systematic expansion
for $\hat{O}$ of the form

\begin{equation} 
\hat{O}=\sum_{n=0}^\infty \hat{O}^{(n)},
\label{eq:expansion}
\end{equation} 
where we have labeled the contributions to $\hat{O}$
by their order $n$ in the small parameter $Q$ of Eq.~(\ref{eq:Q}). This
expansion is partially motivated by the idea of a long-wavelength
limit, where distances smaller than $1/{\Lambda_{\chi \rm{SB}}}$
are not resolved.

Equation~(\ref{eq:expansion}) is an operator statement, and the
nucleon momentum operator $\hat{p}$ appears on the right-hand side.
Thus, one might worry about the unboundedness of this 
quantum-mechanical operator.
However, once we take matrix elements of $\hat{O}$ to construct the
physical amplitude for the process in question the only quantities
which appear are objects like $\langle \psi| \hat{p} |\psi
\rangle$. For weakly-bound systems like the deuteron these 
expectation values are generically small compared to $\Lambda_{\chi {\rm SB}}$.

The procedure to construct $\hat{O}^{(n)}$ begins with writing down
the vertices appearing in the chiral Lagrangian up to order $n$. One
then draws all of the two-body, two-nucleon-irreducible, Feynman
graphs for the process of interest which are of chiral order $Q^n$. The
rules for calculating the chiral order of a particular graph are:

\begin{itemize}
\item Each nucleon propagator scales like $1/Q$;

\item Each loop contributes $Q^4$;

\item Graphs in which both particles participate in the reaction
acquire a factor of $Q^3$;

\item Each pion propagator scales like $1/Q^2$;

\item Each vertex from the $n$th-order piece of the chiral Lagrangian
contributes $Q^n$.  
\end{itemize}

In this way we see that more complicated graphs, involving two-body
mechanisms, and/or higher-order vertices, and/or more loops, are suppressed by
powers of the small parameter $Q$. 

There remains the problem of constructing a deuteron wave function
which is consistent with the operator $\hat{O}$. The
proposal of Weinberg was to construct a $\chi$PT expansion as per
Eq.~(\ref{eq:expansion}) for the $NN$ potential $V$, and then solve
the Schr\"odinger equation to find the deuteron (or other nuclear)
wave function~\cite{Wein}.  Recent calculations have shown that the
$NN$ phase shifts can be understood, and deuteron bound-state static
properties reliably computed, with wave functions derived from
$\chi$PT in this way~\cite{Or96,Ep99,Re99}. The difficulty is that
there is no {\it a priori} reason within $\chi$PT to iterate $V$
to all orders, since the loop corrections which are
responsible for generating the deuteron bound state are, in principle,
higher-order effects in the EFT. In marked contrast to $\chi$PT in the
one-nucleon sector, Weinberg's application of $\chi$PT to nuclei has
power-counting only for the potential, and not for the $NN$ scattering
amplitude. The most notable attempt to remedy this deficiency was made
by Kaplan, Savage, and Wise, in Ref.~\cite{Ka98A}.  I will not discuss
these issues further here, but instead refer to Ref.~\cite{Bd00} for a
survey of the state of play.  In this work I focus on the $\chi$PT
expansion for the operators $\hat{O}$ and use wave functions found in
one of three ways:
\begin{enumerate} 
\item The full chiral wave function of Ref.~\cite{Ep99} (or
equivalently Ref.~\cite{Or96}). This implements Weinberg's proposal to
use the chiral expansion for $V$, and includes terms in the potential up
to chiral order $Q^3$.

\item In the long-wavelength approximation it is valid to employ a
wave function that is the solution to the Schr\"odinger equation with
a potential which is the sum of one-pion exchange and a short-distance
interaction of range $R < 1/m_\pi$~\cite{Pa99,PC99}. The short-distance
potential should be tuned so as to reproduce the deuteron binding
energy $B$ and the asymptotic wave-function normalizations $A_S$ and
$A_D$. Wave functions generated using potentials with different
values of $R$ will then differ from each other only for $r < R$.

\item Similarly, potential-model wave functions, e.~g. the
Nijm93 wave function~\cite{St94}, can be used for
$|\psi \rangle$. Historically these have been the ``wave functions of
choice" for $\chi$PT calculations of reactions on the deuteron. While
not entirely consistent, this ``hybrid" approach
has led to successful calculations of processes including, but not
limited to, $\gamma d \rightarrow \pi^0 d$~\cite{Be97} and $pp
\rightarrow d e^+ \nu_e$~\cite{Pa98B}. Ref.~\cite{vK99} contains a
thorough review of results obtained via this method.
\end{enumerate}

\section{ONE PHOTON: DEUTERON ELECTROMAGNETIC FORM FACTORS}

In this section I will discuss some aspects of electron-deuteron
scattering in $\chi$PT (see~\cite{PC99} for a more thorough
treatment). As is well-known, the interaction of a deuteron with an
electromagnetic current is naturally described in terms of three
nuclear response functions: the charge ($F_C$), magnetic ($F_M$), and
quadrupole ($F_Q$) form factors. These form factors can be extracted
from electron scattering measurements using differential cross-section
and tensor-polarization data. They can also be computed systematically
in $\chi$PT.  The $Q$-expansion of the deuteron charge operator is:

\begin{equation} 
\hat{{\cal Q}}=e\left[1 \qquad + \qquad \left({\rm relativistic \; corrections}
\; - \; \frac{1}{6} q^2 r_N^2\right) \qquad + \qquad O(Q^3)\right],
\label{eq:deuteroncharge} 
\end{equation} 
where the first term is the leading contribution, which comes only
from the charge of a point proton. Corrections to this $O(e)$
result arise at order $e Q^2$ from the finite-size of the nucleon and
from relativistic effects. The latter can be systematically computed
in $\chi$PT and scale as $1/M^2$ ($M$ is the nucleon mass), but I have
not written them explicitly here. At $O(e Q^3)$ $\chi$PT gives
a two-body contribution due to the exchange of pions in
${\cal Q}$. 

This picture of the deuteron charge is similar to that obtained from
systematic $1/M$ expansions in non-relativistic potential models (see
e.g.~\cite{Ar99}), apart from the somewhat unusual expedient of
expanding out the nucleon's charge form factor in powers of $q^2
r_N^2$. Thus, it is reasonable to ask what is gained when $\chi$PT is
employed here.

Firstly, and perhaps most importantly, one gains the ability to
include two-body contributions to $\hat{\cal Q}$ systematically.  In
$\chi$PT such meson-exchange currents can be included as they arise
order-by-order in the $Q$-expansion. This is a critical advantage in
reactions where two-body mechanisms are important, as we shall see in
Section 4.

\begin{figure}[h,t,b,p] 
\centering
{\epsfig{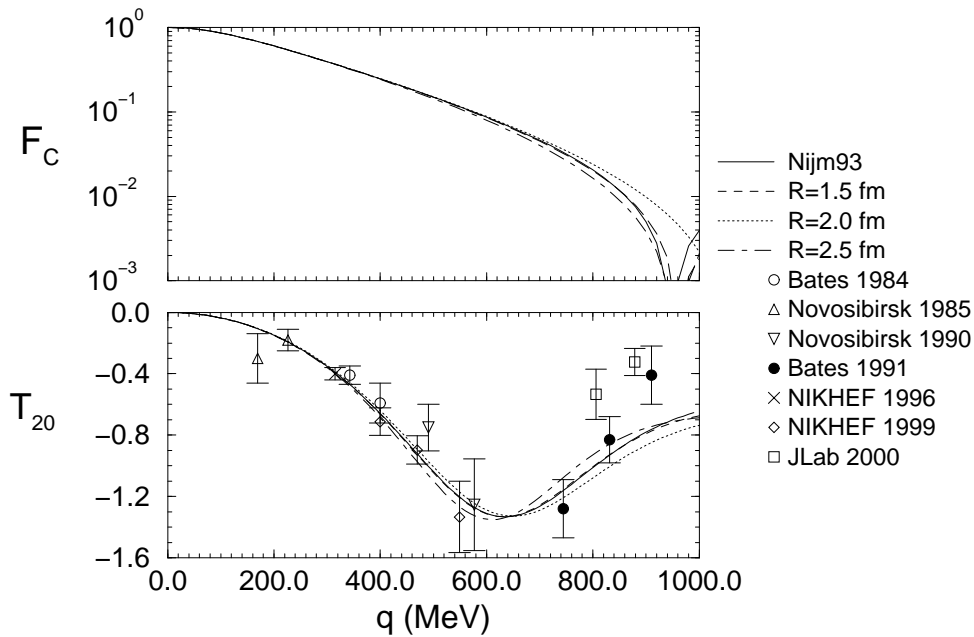}} \caption[]{The upper
panel shows the charge form factor of the deuteron for several
different wave functions, all of which have the same tail, as well as
the same one-pion exchange part. The solid line is the result for the
Nijm93 potential~\cite{St94}. The lower panel depicts the tensor polarization
observable $T_{20}$ for the same four wave functions, compared to data
below $q=1$ GeV. Data are taken from Ref.~\cite{Ab00}.}  
\label{fig-ed}
\end{figure}

Secondly, on an aesthetic level $\chi$PT leads to what might be
considered a more elegant formulation---one where the amount of effort
required to obtain accurate answers is commensurate with the physics
of the long-wavelength limit. In particular, it is straightforward to
compute wave functions and electromagnetic operators consistently, and
many of the issues which arise due to the complicated form of modern
$NN$ potentials do not occur in a $\chi$PT computation of
electron-deuteron scattering. In fact, it is somewhat surprising how
well a very simple picture of the deuteron describes the charge and
magnetic form factors. In the upper panel of Fig.~\ref{fig-ed} I have
displayed the charge form factor computed with the leading-order
charge of Eq.~(\ref{eq:deuteroncharge}) and a variety of wave
functions obtained with the simple OPEP plus short-distance potential
described in the previous section~\cite{PC99}. One sees that the
sensitivity to the physics at distances smaller than $R$ is slight,
appearing only at momentum transfers of order 600 MeV. Furthermore,
for a range of values of $R$ the result for $F_C$ from these simple
wave functions is close to that of the much more sophisticated Nijm93
wave function.

A third role of $\chi$PT is to point to physics not currently included
in potential-model descriptions of electron-deuteron scattering. For
instance, a survey of the results for deuteron static properties found
using the same wave functions employed in the calculations for $F_C$
reveals that the deuteron quadrupole moment, $Q_d$, is particularly
susceptible to short-distance physics. It varies from $0.258 \; {\rm
fm}^2$ when the radius of the short-distance potential, $R$, is $2.5 \;
{\rm fm}$ to $0.269 \; {\rm fm}^2$ when $R=1.5 \; {\rm fm}$. Such
sensitivity of $Q_d$ to short-range dynamics will be no surprise to
those familiar with $NN$ potential models. However, in an EFT it
suggests the presence of a two-nucleon one-quadrupole-photon
counterterm which modifies $Q_d$~\cite{PC99,Ch99}. This is physics
``beyond the nuclear standard model", since such mechanisms are not
included in the usual calculations of $Q_d$. An analysis of the 
counterterm required to shift $Q_d$ to its experimental value
of $0.286 \; {\rm fm}^2$ suggests that it is ``natural", i.e. of a 
size consistent with the scales in the problem. This counterterm,
with its value adjusted to give the experimental value of $Q_d$ for
each different wave function, is easily included in the calculation of
the tensor polarization observable $T_{20}$. The lower panel of
Fig.~\ref{fig-ed} then shows that (a) the residual sensitivity to the
short-distance behaviour of the wave function is very small, and
(b) using only the $O(e)$ charge operator and the $Q_d$
counterterm we can describe almost all of the existing $T_{20}$ data
out to $q=700$ MeV.

\section{TWO PHOTONS: COMPTON SCATTERING ON THE DEUTERON}

In this section I will discuss Compton scattering on the deuteron in
$\chi$PT. (Ref.~\cite{Be99} contains a full treatment.) This reaction
has been the subject of recent experiments~\cite{Lu94,Ho00}, one goal
of which was to extract the electromagnetic polarizabilities of the
neutron.

In the case of the proton these electric and magnetic polarizabilities,
$\alpha$ and $\beta$, have been extracted from Compton
scattering data~\cite{To98}:
\begin{equation}
\alpha_p + \beta_p=13.23 \pm 0.86^{+0.20}_{-0.49} \times 10^{-4} \, {\rm fm}^3;
\qquad
\alpha_p - \beta_p=10.11 \pm 1.74^{+1.22}_{-0.86} \times 10^{-4} \, {\rm fm}^3,
\label{eq:protpolexpt}
\end{equation}
These numbers agree well with the {\it predictions} of $\chi$PT
at leading loop order ($O(e^2 Q)$)~\cite{Br95}:
$\alpha_p=12.2 \times 10^{-4} \, {\rm fm}^3$ and
$\beta_p= 1.2 \times  10^{-4} \, {\rm fm}^3$.
%
%
At this order $\chi$PT also predicts $\alpha_n=\alpha_p$ and
$\beta_n=\beta_p$.  $\alpha_n$ and $\beta_n$ are difficult to extract
from experiments. In particular, their difference is largely
unconstrained by data. $\alpha_n - \beta_n$ will play
a role in Compton scattering on deuterium, and one might hope to
measure it there.  However, if a reliable extraction of this difference
is to be made from $\gamma d$ data a formalism must be employed in
which the contribution of two-body currents to the Compton cross
section is under control.

\begin{figure}[h,t,b,p] \centering
{\epsfig{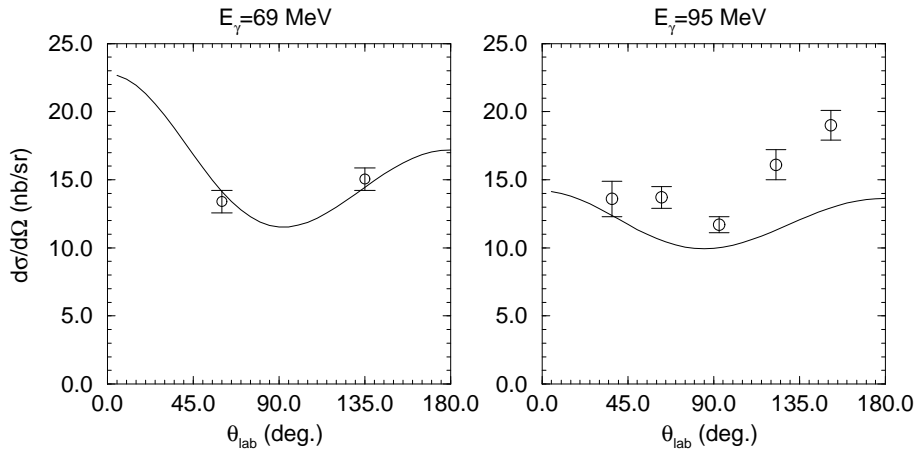}} 
\caption[]{Compton scattering on the deuteron at 69 and 95 MeV in chiral 
perturbation theory at $O(e^2 Q)$ as compared to data from Illinois~\cite{Lu94}
and Saskatoon~\cite{Ho00} respectively.}  
\label{fig-compton} 
\end{figure}

The two-body contributions to $\gamma d$ scattering can be calculated
in $\chi$PT at $O(e^2 Q)$ in the manner described in Section 2. The
resultant graphs are just the two-body analogues of the one-nucleon
graphs which yield the successful $\chi$PT $O(e^2 Q)$ prediction of
$\alpha_p$ and $\beta_p$. By sandwiching these two-body graphs between
the wave-functions of Ref.~\cite{Ep99}, and including the one-body
$O(e^2 Q)$ $\chi$PT $\gamma N$ contribution, we generate the
parameter-free prediction of $\chi$PT at $O(e^2 Q)$ for $\gamma d$
differential cross sections for the Compton scattering process.  Our
result should agree with data over a range of energies from about 50
MeV up to $m_\pi$~\cite{Be99}. The comparison is made in
Fig.~\ref{fig-compton} for photon energies of 69 and 95 MeV. There is
good agreement with the (limited) data at 69 MeV, but our results
disagree with the backward-angle data at 95 MeV. This disagreement
also appears in potential-model calculations of $\gamma d$ scattering,
unless values of $\alpha_n - \beta_n$ very different from $\alpha_p -
\beta_p$ are employed~\cite{LL99}. In fact, comparing such
computations to our $\chi$PT approach to $\gamma d$ scattering
illustrates the utility of $\chi$PT. The calculation of
Ref.~\cite{LL99} is significantly more complicated,
but gives similar results to our Ref.~\cite{Be99}. These complications arise
because the potential-model approach does not exploit the
simplifications offered by the long wavelength of the photons.

\section*{ACKNOWLEDGEMENTS}
I thank Silas Beane, Tom Cohen, Manuel Malheiro, and
Bira van Kolck for enjoyable collaborations on the topics discussed in
this talk. Thanks also to Matt Dorsten for his help in generating Figure
1(b), and to Evgeni Epelbaum and Vincent Stoks for supplying deuteron
wave functions. I am grateful to Silas for his comments on
the manuscript. This research was supported by the U.~S. Department of Energy
(grants DE-FG02-93ER-40762 and DE-FG03-97ER-41014).

\end{document}